\documentclass[journal,10pt,draftclsnofoot,onecolumn]{IEEEtran}
\usepackage[dvipdfm]{graphicx}
\usepackage{comment}
\usepackage[applemac]{inputenx}     
\usepackage{cite}
\usepackage[usenames,dvipsnames]{color}
\usepackage{amssymb}
\usepackage{amsmath}
\usepackage{epsf}
\usepackage{epsfig}
\usepackage{epstopdf}
\usepackage{amsfonts}%
\usepackage{graphicx}%
\usepackage{multicol}%
\usepackage{verbatim}%
\usepackage{enumerate}
\usepackage{amsthm}
\usepackage{eucal}
\usepackage{subfigure}

\begin{document}
\title{Pseudo-Lattice Treatment for Subspace Aligned Interference Signals}

\author{Yahya H. Ezzeldin$^\dagger$ and Karim G. Seddik$^*$\\ [.1in]
 \begin{tabular}{c}
$^\dagger$Department of Electrical Engineering, Alexandria University, Alexandria, Egypt \\
$^*$Electronics Engineering Department, American University in Cairo, AUC Avenue, New Cairo, Egypt.\\
email: yahya.ezzeldin@ieee.org, kseddik@aucegypt.edu
\end{tabular}
}

\maketitle

\begin{abstract}
For multi-input multi-output (MIMO) $K$-user interference networks, we propose the use of a channel transformation technique for joint detection of the useful and interference signals in an interference alignment scenario. We coin our detection technique as \textit{``pseudo-lattice treatment"} and show that applying our technique, we can alleviate limitations facing Lattice Interference Alignment (L-IA). We show that for a 3-user interference network, two of the users can have their interference aligned in lattice structure through precoding. For the remaining user, performance gains in decoding subspace interference aligned signals at the receiver are achieved using our channel transformation technique. Our ``pseudo-lattice" technique can also be applied at all users in case of Subspace Interference Alignment (S-IA). We investigate different solutions for applying channel transformation at the third receiver and evaluate performance for these techniques. Simulations are conducted to show the performance gain in using our pseudo-lattice method over other decoding techniques using different modulation schemes.
\end{abstract}

\section{Introduction}
Management of interference is one of the key challenges for spectrum sharing in wireless communications. Different techniques for managing interference among users sharing the same frequency band have been developed in the literature. As characterization of available communication resources, the number of degrees of freedom (DoF) are used to represent the interference-free dimensions in a communication network which can be shared between the interfering users. For two-user interference channel, different techniques can be used to achieve 1/2 of the DoF per user, through time division or spatial multiplexing, etc.

If we scale the problem to a $K$-user interference network, the pre-mentioned solutions can achieve as much as $1/K$ of the DoF per user. In \cite{Jaafar,Jaafar2}, \textit{interference alignment}, a strategy for handling interference in multiuser interference networks is introduced that is shown to achieve $1/2$ DoF per user independent of the number of users in the network. The technique precodes signals transmitted from each of the $K$ transmitters such that for each receiver, interference from the $K-1$ users is confined in half the received signal space. Therefore, the problem returns to being a two-user interference network where one transmitter sends the desired signal using half the DoF and a virtual interfering node transmits the collective interference from the $K-1$ transmitters over the remaining DoF.

In the paper, we focus on the DoF exploited by having $N > 1$ antennas at each of the $K$ transmitter and receiver pairs. The achievable DoF in this case is $N/2$ (given that $N$ is even). Although interference is limited to a subspace at each receiver, the use of ML detectors is complex specifically as the signal dimensions increase with the number of antennas.

In \cite{L-IA1}, \textit{Choi} shows that the use of low complexity Lattice Reduction (LR) decoders \cite{LR-1,LR-2} at the receivers can substantially improve the decoding performance while maintaining low complexity. As a limitation for the technique mentioned, only two users in a 3-user network are able to benefit from lattice interference alignment while the remaining user will have to treat the interference as subspace aligned.

In this paper, we define a \textit{pseudo-lattice treatment} that allows usage of LR-based detectors in scenarios where only subspace alignment is possible. Our detector can be used in subspace interference alignment (S-IA) for all users or in Lattice Interference Alignment (L-IA) to overcome the pre-mentioned implementation limitation for all users. Note that, the pseudo-lattice detector requires no further coordination by the transmission than what S-IA or L-IA require; the channel transformation to enforce lattice structured interference is performed by the receiver.

The remainder of the paper is organized as follows: In Section \ref{sysmodel_sec}, we present the model for our MIMO multiuser interference channel. In Section \ref{framework_sec}, we discuss basic Interference Alignment schemes and how our proposed treatment differs from the discussed schemes. Finally, in Section \ref{analysis_sec}, we present a mathematical analysis for the performance of the receiver in terms of decoding error probability when our framework is implemented. In the same section, we discuss different examples of how the channel transformation treatment can be implemented and how the system performs under these different implementations. Finally, in Section \ref{simulations_sec} we present simulation results and then follow with conclusion on the work presented in the paper.

\textit{Notations:} Throughout the paper, we use $\mathbb{Z}$ to refer to the real integers and $\mathbb{C}$ to refer to the complex numbers. We represent vectors with bold lower cases such as $\mathbf{x} \in \mathbb{C}^N$. Matrices are represented by bold upper cases such as $\mathbf{A} \in \mathbb{C}^{M \times N}$. The superscripts $\mathsf{T}$ and $\mathsf{H}$ denote the transpose and Hermitian transpose of a matrix, respectively. The $\ell_2$-norm of vector $\mathbf{a}$ is denoted by $\Vert \mathbf{a} \Vert$. $\mathbf{A}^\dagger$ and $\Vert \mathbf{A} \Vert_{}$ denote the pseudo-inverse and Frobenius norm of matrix $\mathbf{A}$, respectively.

\section{System Model}\label{sysmodel_sec}
\begin{figure}
	\centering
	\subfigure[System Model for a $K$-user Interference Alignment network.]{
	\centering	
	\includegraphics[width=0.47\textwidth]{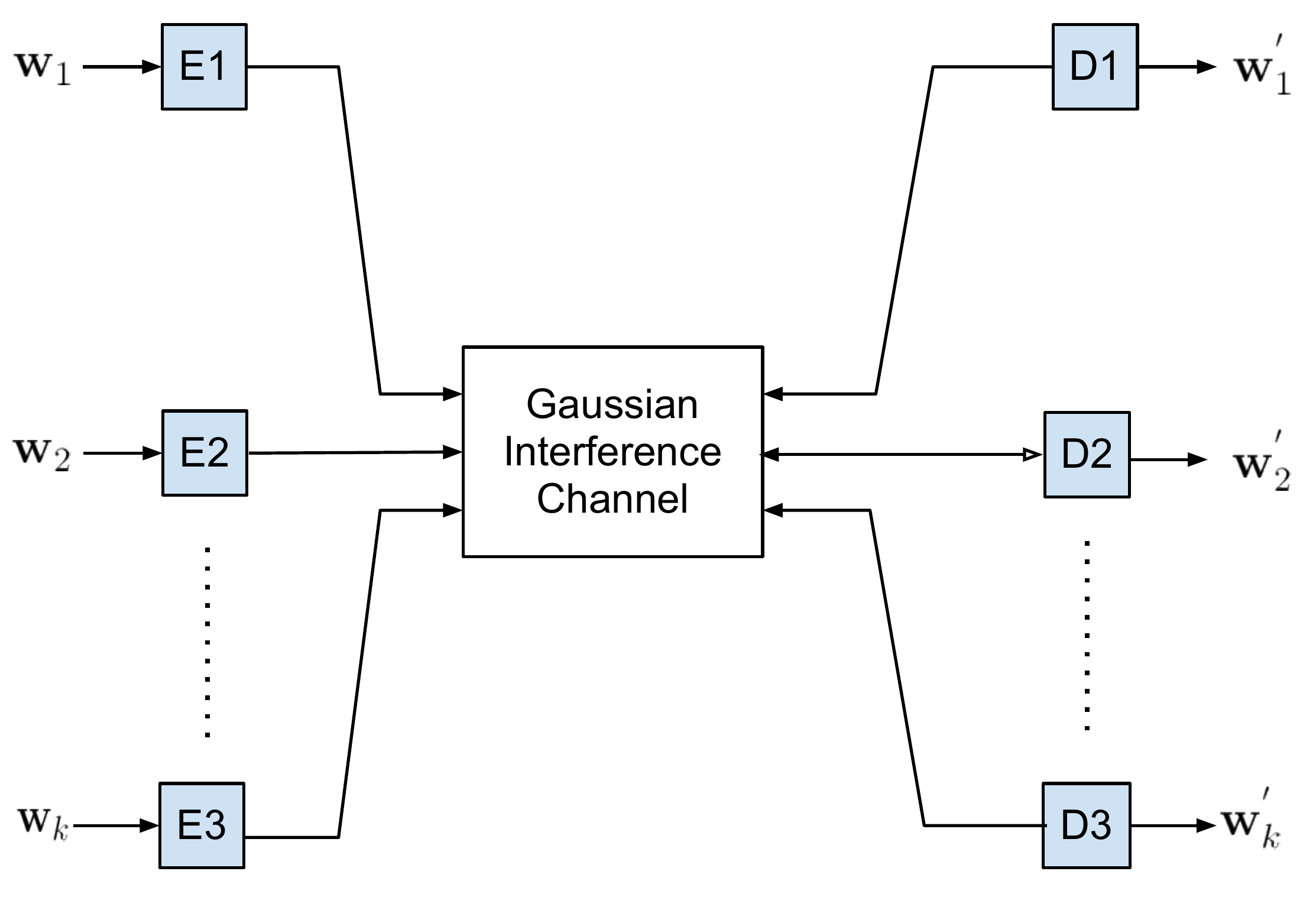}
	\label{fig:model}
}
	\subfigure[Lattice $\mathbf{\Lambda}_i$ at user $i$ with desired signal aligned in the direction of the basis $b_1$interference aligned in the direction of the basis $b_2$,  the number of antennas ($N = 2$).]{
	\includegraphics[width=0.47\textwidth]{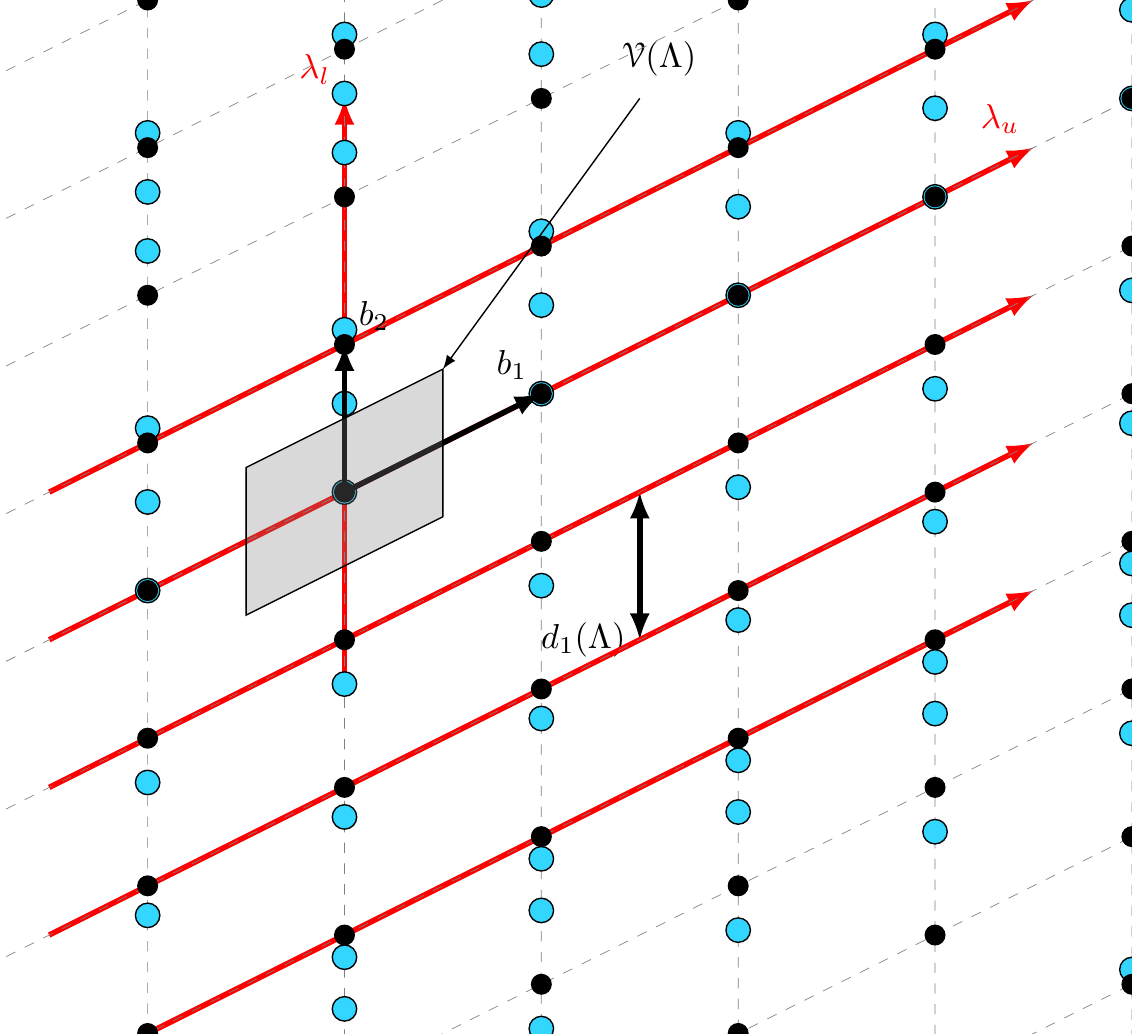}
	\label{figure:lattice}
}
	\caption{System Model}
\end{figure}

We consider $K$ transmitter-receiver pairs each equipped with an even number of antennas, $N > 1$. Each of the $K$ transmitters generates a message signal $\mathbf{w}_i \in \mathbb{C}^{N/2}$ to be transmitted through the interference channel and intended for receiver $i$. The messages are precoded before transmission using $\mathbf{B}_i \in \mathbb{C}^{N \times N/2}$ and fed to the Gaussian Interference channel as seen in Figure \ref{fig:model}. The received signal at receiver $i$ is:
\begin{equation}\label{sys_model}
\mathbf{r}_i = \sum_{j = 1}^{K} \mathbf{H}_{i,j} \mathbf{B}_{j} \mathbf{w}_j + \mathbf{z}_i\ ,
\end{equation}
where $\mathbf{H}_{i,j}$ denotes the $N \times N$ channel matrix between transmitter $j$ and receiver $i$. The channel coefficients in $\mathbf{H}_{i,j}$ are i.i.d. complex Gaussian channels with zero mean and variance $\sigma^2$. $\mathbf{z}_i$ is the background AWGN noise at receiver $i$, such that $\mathbf{z}_j \sim \mathcal{CN}\ (0, \sigma_z^2$). The transmitted message signals $\mathbf{w}_i$ are drawn from a finite lattice constellation (for example: QAM constellations) as in Figure \ref{figure:lattice}. The precoded transmitted signal is of normalized power $P_{\rm s}$ such that $\mathbb{E}\lbrace\;\Vert\mathbf{B}_i \mathbf{w}_i \Vert_2^2\;\rbrace = P_{\rm s}$ for all $i$.
Channel coefficients $\mathbf{H}_{i,j}$ are assumed to be perfectly known at the transmitters and receivers.

\subsection*{Transmitter and Receiver Lattice Structure:}
\noindent In our model, we consider signals transmitted that have a lattice structure. Lattice $\mathcal{L}(\mathbf{G})$ represents the lattice generated using the \textit{generator matrix} $\mathbf{G}$, i.e: 
\begin{equation}
\mathcal{L}(\mathbf{G}) = \lbrace \mathbf{G}\mathbf{x}\ \vert\ \mathbf{x} \in \mathbb{Z}^{N/2} + j\mathbb{Z}^{N/2} \rbrace = \Lambda_G.
\end{equation}

If we consider lattice $\mathcal{L}(\mathbf{G}) \in \mathbb{C}^{N/2}$ is transmitted from each of $K$ users, then the received signal is a superposition of several lattices. Generically, the received signal has no lattice structure as the channel coefficients can take non-integer values. Assuming that Interference Alignment is satisfied (conditions shall be discussed in Section \ref{framework_sec}), the desired signal will span the first $N/2$ dimensions of the decoded signal space. We denote the first $N/2$ dimensions of the decoded lattice as $\lambda_{i,u} \in \Lambda_{i,u}$ where:
\begin{equation}
\Lambda_{i,u} = \mathcal{L}(\mathbf{H}_{ii} \mathbf{B}_i).
\end{equation}
Now assuming that lattice interference alignment is achieved, the remainder of decoded signal vector, denoted as $\lambda_{i,l}$ takes a lattice structure $\mathbf{\Lambda}_{i,l}$ as well. Therefore the decoded signal can be represented as $\lambda = \left[ \lambda_{i,u}\ \ \lambda_{i,l}\right]^\mathsf{T}$ where $\lambda_{i,u} \in \mathbf{\Lambda}_{i,u}$ and $\lambda_{i,l} \in \mathbf{\Lambda}_{i,l}$. This structure can be seen in Figure \ref{figure:lattice} where the lattice is defined with signals along the dimension of the basis of $\mathbf{\Lambda}_{i,u}$ and interfering signals along the basis of $\mathbf{\Lambda}_{i,l}$.

For a lattice $\Lambda$ we define \textit{Voronoi Region}, $\mathcal{V}_{\lambda}(\Lambda)$, as the set of points that are closer to $\lambda$ than any point in the lattice $\Lambda$. The points in this region would be mapped to the center of the Voronoi region by a maximum likelihood (ML) decoder. All LR algorithms try to achieve the same incentive with lower complexity.

In the following section, we discuss the different techniques for lattice interference alignment that lead up to our pseudo-lattice treatment.
\section{Interference Alignment \& Proposed Framework}\label{framework_sec}
At receiver $i$, the signal in \eqref{sys_model} contains the desired message $\mathbf{w}_i$ as well as interference components $\sum_{j \neq i} \mathbf{H}_{i,j} \mathbf{B}_{j} \mathbf{w}_j$. Using interference alignment\cite{Jaafar,Jaafar2}, we can confine the interference to an $N/2$ subspace. The $N$ degrees of freedom are now sufficient to decode the desired signal $\mathbf{w}_i$ which resides in an $N/2$ space. The nominal interference alignment setting is to decode when the interference in aligned in a subspace of the signal space. 
\subsection{Subspace Interference Alignment (S-IA)}
In Subspace Interference Alignment, the transmitters precoders confine the interference received at the receiver to a signal subspace of $N/2$ dimensions. For a $K$-user system, the condition for aligning interference at receiver 1 can be expressed as:
\begin{equation}
\text{span}(\mathbf{H}_{1,2} \mathbf{B}_{2}) = \text{span}(\mathbf{H}_{1,3} \mathbf{B}_{3}) = \cdots = \text{span}(\mathbf{H}_{1,K} \mathbf{B}_{K})\ .
\end{equation}
For subspace interference alignment, decoding of the desired signal can be performed using Zero-Forcing detector (ZF).


\subsection{Lattice Interference Alignment (L-IA)}
In \cite{L-IA1}, a framework for aligning lattice structures instead of spaces have been introduced for a $3$-user system. In this framework, the users aim to match their respective lattice generating functions such that: 
\begin{equation}\label{lattice_eq1}
\mathcal{L}(\mathbf{H}_{1,2} \mathbf{B}_{2}) = \mathcal{L}(\mathbf{H}_{1,3} \mathbf{B}_{3}) 
\end{equation}
\begin{equation}\label{lattice_eq2}
\mathcal{L}(\mathbf{H}_{2,1} \mathbf{B}_{1}) = \mathcal{L}(\mathbf{H}_{2,3} \mathbf{B}_{3})
\end{equation}
\begin{equation}\label{lattice_eq3}
\mathcal{L}(\mathbf{H}_{3,1} \mathbf{B}_{1}) = \mathcal{L}(\mathbf{H}_{3,2} \mathbf{B}_{2})
\end{equation}
where $\mathcal{L}(\mathbf{G})$ represents the lattice generated using the \textit{generator matrix} G, (i.e: $\mathcal{L}(\mathbf{G}) = \lbrace \mathbf{G}\mathbf{x}\ \vert\ \mathbf{x} \in \mathbb{Z}^{N/2} + j\mathbb{Z}^{N/2} \rbrace$).
As Choi discusses in \cite{L-IA1}, these conditions cannot be met for interference at each of the receivers and we are only able to achieve alignment for two of the receivers. For the remaining receiver, the condition is reverted to nominal subspace interference alignment and ZF can be applied to decode the desired signal for that receiver. We refer the reader to \cite{L-IA1} for further comparison between the performance of the two methods.


\subsection{Pseudo-Lattice Treatment}
In this paper, we consider lattice treatment for subspace aligned interference signals. A virtue of having a lattice structure for the signal to be decoded is that we can apply low complexity lattice reduction algorithms to decode the desired message $\mathbf{w}_i$ and the effective interference aligned signals.

Without loss of generality, we will discuss a 3-user interference system. The generalization to a $K$-user system flows directly as the operation discussed below can be applied for an arbitrary number of interfering streams. If we consider that lattice interference alignment has been achieved at user 2 and user 3 while subspace alignment is achieved at user 1, then the precoders are designed such that equations \eqref{lattice_eq2} and \eqref{lattice_eq3} are satisfied in addition to:
\begin{equation}\label{space_IA_1}
\text{span}(\mathbf{H}_{1,2} \mathbf{B}_{2}) = \text{span}(\mathbf{H}_{1,3} \mathbf{B}_{3}).
\end{equation}

The interference from user 2 and user 3 is now aligned in a subspace that can be expressed as the column space of $\mathbf{H}_{1,2} \mathbf{B}_{2}$ or $\mathbf{H}_{1,3} \mathbf{B}_{3}$. Although the interference is subspace aligned, it does not exhibit a lattice structure even though individual interference sources still retain their lattice structure. We denote $\Lambda_2$ and $\Lambda_3$ as the lattice received as interference from users 2 and 3, respectively. $\Lambda_2$ and $\Lambda_3$ can be expressed as:
\begin{equation*}
\Lambda_2 = \mathcal{L}(\mathbf{H}_{1,2} \mathbf{B}_{2})
\end{equation*}
\begin{equation}\label{general_lattices}
\Lambda_3 = \mathcal{L}(\mathbf{H}_{1,3} \mathbf{B}_{3}).
\end{equation}
Although $\Lambda_2$ and $\Lambda_3$ are lattices, $\Lambda_2 + \Lambda_3$ is not generically a lattice. This is because $\Lambda_2 \nsubseteq \Lambda_3$ and $\Lambda_3 \nsubseteq \Lambda_2$, therefore there is no ordered structure in the received signal. We can rewrite the expressions in \eqref{general_lattices} in terms of a common factor as follows:
\begin{equation*}
\Lambda_2 = \mathcal{L}(\mathbf{H}_{c}\mathbf{D}_{2} )
\end{equation*}
\begin{equation}\label{mapping_channels}
\Lambda_3 = \mathcal{L}(\mathbf{H}_{c}\mathbf{D}_{3} ).
\end{equation} 

The expressions in \eqref{mapping_channels} represent a redefinition of the interference subspace in terms of new basis $\mathbf{H}_c \mathbf{D}_j$, such that: $\mathbf{D}_{j} = \mathbf{H}_c^\dagger\ \mathbf{H}_{1,j} \mathbf{B}_{j}$ $\in \mathbb{C}^{N \times N/2}$. For $\Lambda_2$ and $\Lambda_3$ to be part of a larger lattice structure $\mathcal{L}(\mathbf{H}_{\rm c})$, $\mathbf{D}_2$ and $\mathbf{D}_3$ need to belong to  $\mathbb{Z}^{N \times N/2} + j\mathbb{Z}^{N \times N/2}$). As explained earlier, finding $\mathbf{D}_2$ and $\mathbf{D}_3$ with complex integer terms such that $\Lambda_2 \nsubseteq \Lambda_3$ or  $\Lambda_3 \nsubseteq \Lambda_2$ is not possible in most cases as pre-mentioned. The matrices $\mathbf{D}_2$ and $\mathbf{D}_3$ are usually in complex terms $\mathbb{C}^{N \times N/2}$ rather than complex integers. The received signal space therefore does not exhibit a lattice structure as can be seen in Figure \ref{figure:lattice}.

We define $\varphi : \mathbb{C}^{N \times N/2} \longrightarrow \mathbb{Z}^{N \times N/2} + j\mathbb{Z}^{N \times N/2}$ as a surjective mapping function than maps a matrix $\mathbf{A}$ to the nearest complex integer matrix $\varphi(\mathbf{A})$. The receiver, can find groups of matrices $\lbrace \mathbf{H}_c , \mathbf{D}_2 , \mathbf{D}_3 \rbrace$ such that 
\begin{equation}\label{lattice_approx}
\mathcal{L}(\ \mathbf{H}_{c}\;\varphi(\mathbf{D}_{2})\ )\ ,\ \mathcal{L}(\ \mathbf{H}_{c}\;\varphi(\mathbf{D}_{3})\ ) \ \subseteq\ \mathcal{L}(\ \mathbf{H}_{c}) 
\end{equation}
The approximation in \eqref{lattice_approx} introduces approximation noise to the received signal. We can rewrite the interference signal from \eqref{sys_model} as
\vspace{-0.1in}
\begin{eqnarray}\label{sys_model_approx}
\nonumber \mathbf{r}_i &=&\vspace{-0.5in}\sum_{j = 1}^{K} \mathbf{H}_{i,j} \mathbf{B}_{j} \mathbf{w}_j + \mathbf{z}_i \;=\; \vspace{-0.1in} \mathbf{H}_{i,i} \mathbf{B}_{i} \mathbf{w}_i + \sum_{j \neq i}^{K} \mathbf{H}_{c} \mathbf{D}_j \mathbf{w}_j + \mathbf{z}_i\\
&=&\vspace{-0.1in} \underbrace{\mathbf{H}_{i,i} \mathbf{B}_{i} \mathbf{w}_i}_\text{Desired Signal} + \underbrace{\sum_{j \neq i}^{K}  \mathbf{H}_{c}\;\varphi(\mathbf{D}_j) \mathbf{w}_j}_\text{Lattice Interference} \;+\; \underbrace{\underbrace{\sum_{j \neq i}^{K} \mathbf{H}_{c}\;(\mathbf{D}_j - \varphi(\mathbf{D}_j) \mathbf{w}_j}_\text{Approximation noise} + \mathbf{z}_i}_\text{Effective noise ($\mathbf{n}_i$)},
\nonumber
\end{eqnarray}
where the effective noise $\mathbf{n}_i$ is the total noise seen by the LR decoder due to the background noise $\mathbf{z}_i$ and the noise due channel approximation. Note that, this is similar to the concept of lattice decoding of linear combination in Compute-And-Forward (C\&F) relaying \cite{CF-relay1} and Physical Network Coding in \cite{Lattice-PNC}.
For a 3-user system, the incentive is to find the matrix tuple $\lbrace \mathbf{H}_c , \mathbf{D}_2 , \mathbf{D}_3 \rbrace$, that minimizes the difference between the effective noise (\textbf{$\mathbf{n}_i$}) and background noise ($\mathbf{z}_i$), i.e.:
\begin{equation*}
\lbrace \mathbf{H}_c , \mathbf{D}_2 , \mathbf{D}_3 \rbrace = \min_{\mathbf{H}_c , \mathbf{D}_2 , \mathbf{D}_3} \sum_{j=2}^3\Vert \mathbf{H}_{c}\;(\mathbf{D}_j - \varphi(\mathbf{D}_j)\ )\Vert_{F}^2
\end{equation*}
\hspace{0.3in}such that:
\begin{equation} \label{min_tuple}
\mathbf{H}_{c}\mathbf{D}_{j} = \mathbf{H}_{1,j} \mathbf{B}_{j}.
\end{equation}
This is a receiver incentive therefore there is no disruption to lattice interference alignment at other receivers as the precoding matrices are still calculated to satisfy \eqref{lattice_eq2} , \eqref{lattice_eq3} and \eqref{space_IA_1}. 

Note that this lattice approximation is done to the interfering signals and therefore the approximation noise will mainly affect the decoding of the interfering signals which are undesired signals and their decoding error does not contribute to the receiver error rate. 

\subsection*{\underline{Choosing the matrix $\mathbf{H}_{c}$}: }
\noindent As mentioned in the previous subsection, for a $3$-user system, the target is to find the tuple $\lbrace \mathbf{H}_c , \mathbf{D}_2 , \mathbf{D}_3 \rbrace$ that satisfies \eqref{min_tuple}.
The Frobenius norm $\Vert A \Vert_{F}^2$ where $A \in \mathbb{C}^{N \times N/2}$ is the regular vector Euclidean norm when $A$ is viewed as $\frac{N^2}{2}$-dimensional vectors. 
Therefore, the problem in \eqref{min_tuple} maps into a point to a sum of Euclidean norms problem. Several algorithms for minimizing sum of norms are present in the literature \cite{min_Eucl1,min_Eucl2,min_Eucl3	}. In our evaluation of the algorithm, we will be working with simple realization of $\mathbf{H}_{\rm c}$ that limits the space to $\mathbf{H}_{\rm c} \in \lbrace \mathbf{H}_{1,2} \mathbf{B}_{2}\ ,\ \mathbf{H}_{1,3} \mathbf{B}_{3} \rbrace$.
The constitution of $\mathbf{D}_2$ and $\mathbf{D}_2$ changes depending on the chosen $\mathbf{H}_{\rm c}$. For example if we choose $\mathbf{H}_{\rm c} = \mathbf{H}_{1,2} \mathbf{B}_{2}$, then $\mathbf{D}_2$ = $\mathbf{I}$ and $\mathbf{D}_3$ = $\mathbf{H}_{\rm c}^\dagger \mathbf{H}_{1,3} \mathbf{B}_{3}$. Conversely, if we choose choose $\mathbf{H}_{\rm c} = \mathbf{H}_{1,3} \mathbf{B}_{3}$ then $\mathbf{D}_2$ = $\mathbf{H}_{\rm c}^\dagger \mathbf{H}_{1,2} \mathbf{B}_{2}$ and $\mathbf{D}_3$ = $\mathbf{I}$. 
 
\section{Mathematical Analysis}\label{analysis_sec}

The effective noise $\mathbf{n}_i$ denoted in the previous section is not necessarily Gaussian and as a result the analysis becomes non-trivial. To overcome the analysis complexity, we will assume that the shaping region of the higher dimension lattice $\Lambda = \mathcal{L}(\ [\mathbf{H}_{i,i}\ \mathbf{H}_{\rm c}]\ )$ is a rotated hypercube in the complex domain $\mathbb{C}^N$, i.e.,  the Voroni region of $\Lambda$ can be denoted as:
\begin{equation}\label{hypercube}
\mathcal{V}(\Lambda) = \gamma\mathbf{U}\mathcal{H}
\end{equation}
where $\gamma > 0$ is a scalar, $\mathbf{U}$ is a unitary matrix and $\mathcal{H}$ is a unit hypercube in $\mathbb{C}^N$. The assumption of hypercube shaping\cite{shaping1} simplifies the performance analysis, however there is no shaping gain from the unshaped rectangular grid lattice \cite{shaping2,shaping3}.

We define $d(\Lambda)$ as the minimum distance between the two lattice points in $\Lambda$ as follows:
\begin{equation}\label{distance}
d(\Lambda) = \min\lbrace \Vert \lambda_1 - \lambda_2\Vert_2\ \vert\ (\lambda_1\ , \lambda_2) \in \Lambda, \lambda_1 \neq \lambda_2 \rbrace
\end{equation}
For the received lattice $\Lambda = \mathcal{L}(\ [\mathbf{H}_{i,i}\ \mathbf{H}_{\rm c}]\ )$, the minimum distance represents the minimum $\ell_2$-norm of the columns of matrix $[\mathbf{H}_{i,i}\ \mathbf{H}_{\rm c}]$. We also define $E_\varphi$ as the accumulated Frobenious norm of induced noise due to channel approximation of all $K-1$ interference channels using \eqref{min_tuple}:
\begin{equation}\label{error_all}
E_{\varphi}= \sum_{j=1,j\neq i}^{K}\Vert \mathbf{H}_{\rm c}(\mathbf{D}_j - \varphi(\mathbf{D}_j))\Vert_{F}^2\ .
\end{equation}
Now, given our assumption in \eqref{hypercube} and our definitions in \eqref{distance} and \eqref{error_all}, the probability of error ($P_e$) can be approximated as:
\begin{equation}\label{P_e}
P_e \approx \mathbb{E}\bigg\lbrace \mathcal{K}(\Lambda) \exp \bigg(-\dfrac{d^2(\Lambda)}{4\sigma^2_z \big(1 + SNR  (E_{\varphi})\ \  \big)} \bigg)\bigg\rbrace.
\end{equation}
The expression \eqref{P_e} is based on Theorem 7 in \cite{CF-relay1} where $\mathcal{K}(\Lambda)$ represents the number of directions where the distance is the minimum $d(\Lambda)$. i.e.: $\mathcal{K}(\Lambda)$ represents the number of columns of $[\mathbf{H}_{i,i}\ \mathbf{H}_{\rm c}]$ whose $\ell_2$-norm is equal to $d(\Lambda)$. $SNR$ is the ratio between the transmission energy and the noise variance, $SNR = P_{\rm s} / \sigma_z^2$.
For interference alignment, it is worth noting that at receiver $i$, we are only interested in the desired signal space which occupies the first $N/2$ dimensions of the signal space, $\Lambda_{u}$. Therefore in our performance analysis, we will focus on the detection error for signals in the desired signal subspace. We define a new parameter, $d_1(\Lambda)$, as the minimum distance along the lattice basis where the desired signal changes. This can expressed as:
\begin{equation}
d_1(\Lambda) = \min\lbrace \Vert \lambda_1 - \lambda_2\Vert_2\ \vert\ (\lambda_1\ , \lambda_2) \in \Lambda, \lambda_{1,u} \neq \lambda_{2,u} \rbrace,
\end{equation}
where $\lambda_{1,u}$ represent the components in the first $N/2$ dimensions of $\Lambda$ and the remaining $N/2$ components are denoted as $\lambda_{1,l}$. $d_1(\Lambda)$ can be discerned from $\mathbf{H}_{i,i}$. The columns of the matrix $\mathbf{H}_{i,i}$ represent the first $N/2$ basis of the lattice $\Lambda$ and as a result, they represent the directions in which the desired signal component of the lattice changes as seen in Figure \ref{figure:lattice}. We can therefore denote $d_1(\Lambda)$ as the minimum $\ell_2$-norm of the columns of $\mathbf{H}_{i,i}$. The expression in \eqref{P_e} can now be written as:
\begin{equation}\label{P_e2}
P_e\approx \mathbb{E}\bigg\lbrace\mathcal{K}(\Lambda) \exp\bigg(-\dfrac{d_1^2(\Lambda)}{4\sigma^2_z \big(1 + SNR \Vert \mathbf{H}_{c}\Vert_{F}^2 E_{\varphi} \big)} \bigg)\bigg\rbrace .
\end{equation}

\subsection*{\textbf{Performance Bound:}}
The performance of the detector in \eqref{P_e2} depends on the tuple $\lbrace \mathbf{H}_c , \mathbf{D}_2 , \mathbf{D}_3 \rbrace$ as well as the SNR condition of the channel. In low SNR conditions, the effective noise at the receiver is dominated by the background noise. Therefore the expression $P_e$ in \eqref{P_e2} is lower-bounded by:
\begin{equation}\label{Pe_ML}
P_e \geq \mathcal{K}(\Lambda) \exp\bigg(\hspace{-0.06in}-\hspace{-0.02in}\dfrac{d_1^2(\Lambda)}{4\sigma^2_z} \bigg).
\end{equation}
The expression \eqref{Pe_ML} is the performance of the ML receiver in an AWGN channel. 

In high SNR conditions, the effective noise is bounded by the channel approximation noise. Therefore the expression \eqref{P_e2} can be lower bounded in high SNR conditions as:
\begin{equation}\label{Pe_Floor}
P_e \geq \mathcal{K}(\Lambda) \exp\bigg(\hspace{-0.06in}-\hspace{-0.02in}\dfrac{d_1^2(\Lambda)}{4\ P_{\rm s}\ E_{\varphi} } \bigg).
\end{equation}
The expression \eqref{Pe_Floor} shows that the performance of the system experiences an error floor as SNR increases. This can be seen through the simulations in the following section.

\begin{figure}
\centering
	\subfigure[$N = 4$, QPSK.]{
	\centering
	\includegraphics[width=0.45\textwidth]{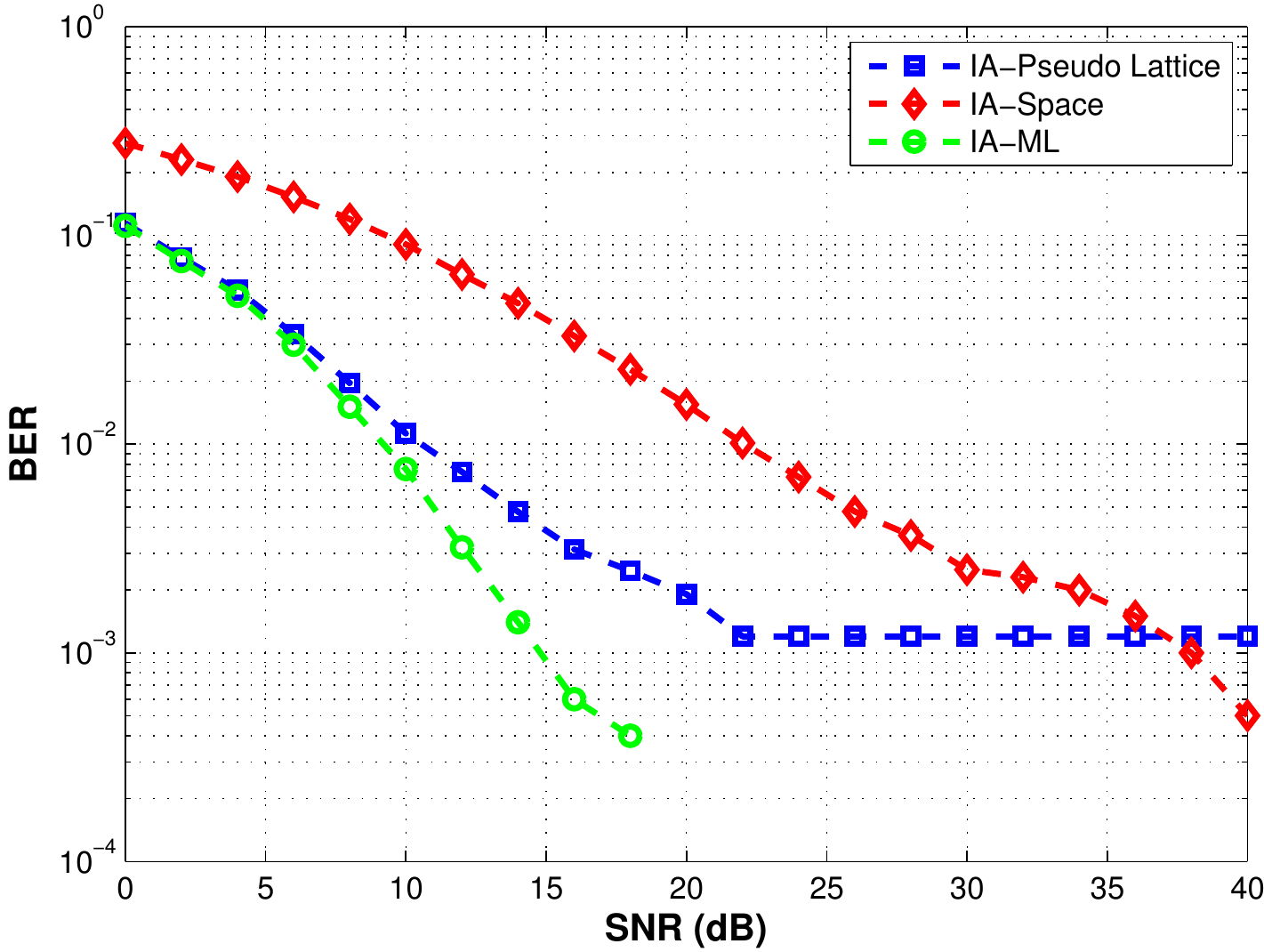}
	\label{figure:BER}
}
	\subfigure[$N = 4$, 16QAM.]{
	\centering
	\includegraphics[width=0.45\textwidth]{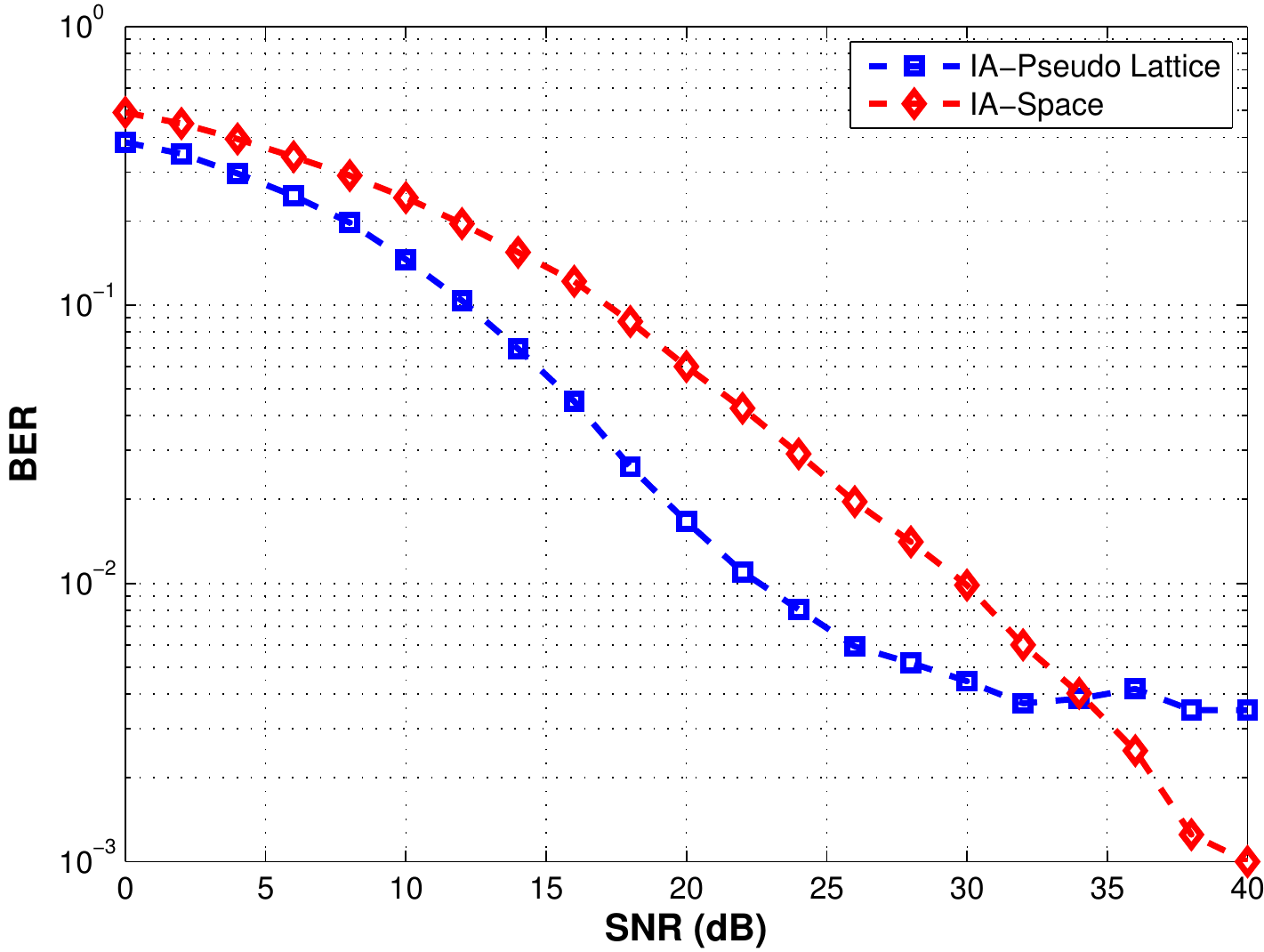}
	\label{figure:BER_QAM}
}			\vspace{-0.1in}
		\caption{BER Vs. SNR for the pseudo-lattice detector with S-IA precoders. Number of antennas $N$ = 4, the channel variance $\sigma_h^2 = 1$ and the noise variance $\sigma_z^2 = 1$}
					\vspace{-0.2in}
\end{figure}
\section{Simulation Results}\label{simulations_sec}
For simulation of the proposed system, Quadrature Phase Shift Keying (QPSK) modulation with gray coding is used. A 3-user system is simulated and it is assumed that the elements of the interference channels $\mathbf{H}_{i,j}$ are i.i.d. with complex Gaussian distribution with circularly symmetric zero mean and unity variance. We assume that the power is normalized at each of the transmitters where $P_{\rm s} = 1$.

For comparison, our pseudo-lattice detector at user 1 is compared with the ZF receiver used in \cite{L-IA1} and benchmarked against the ML-detector. For the remaining two users, interference alignment over a lattice structure is achieved. The results in case of lattice interference alignment is thoroughly covered by \cite{L-IA1}. The matrix $\mathbf{H}_{c}$ is chosen such that $\mathbf{H}_{c} \in \lbrace \mathbf{H}_{1,2} \mathbf{B}_{2}\ ,\ \mathbf{H}_{1,3} \mathbf{B}_{3} \rbrace$ and $\mathbf{D}_{\rm j} = \mathbf{H}_{\rm c}^\dagger\mathbf{H}_{1,j} \mathbf{B}_{j}$. The choice of $\mathbf{H}_{c}$ is based on which of the two realizations minimizes the expression in \eqref{min_tuple}.

Figure \ref{figure:BER} shows the bit error rate (BER) when the number of antennas $N= 4$. We can see from the figure that the proposed pseudo-lattice detector provides better performance for subspace interference alignment than the receiver used in \cite{L-IA1,linear-receiver} in low and medium SNR conditions. Our proposed detector exhibits performance close to the ML-receiver in these SNR ranges. The pseudo-lattice detector approximates a common basis for the the interference subspace such that the whole signal space can now fit into a lattice structure. While, this provides substantial gains in performance in low SNR conditions, the approximation noise becomes more significant as SNR improves. Therefore, Figure \ref{figure:BER} shows an error floor in high SNR ranges for the pseudo-lattice detector.

In Figure \ref{figure:BER_QAM}, we repeat the simulation using 16QAM modulation. We note that for this scenario, the ML performance is not included as given the 3-user network with 4 antennas, exhaustive search would have to go through $16^6$ combinations for ML detection. The performance of the pseudo-lattice detector is better than ZF decoder in low and medium SNR ranges. The ZF receiver first outperforms our detector at 35 dB. At a bit error rate of $10^{-2}$, we can observe a gain of about 7 dB for our proposed detector as compared to the ZF detector.

\begin{figure}
\centering
	\includegraphics[width=0.45\textwidth]{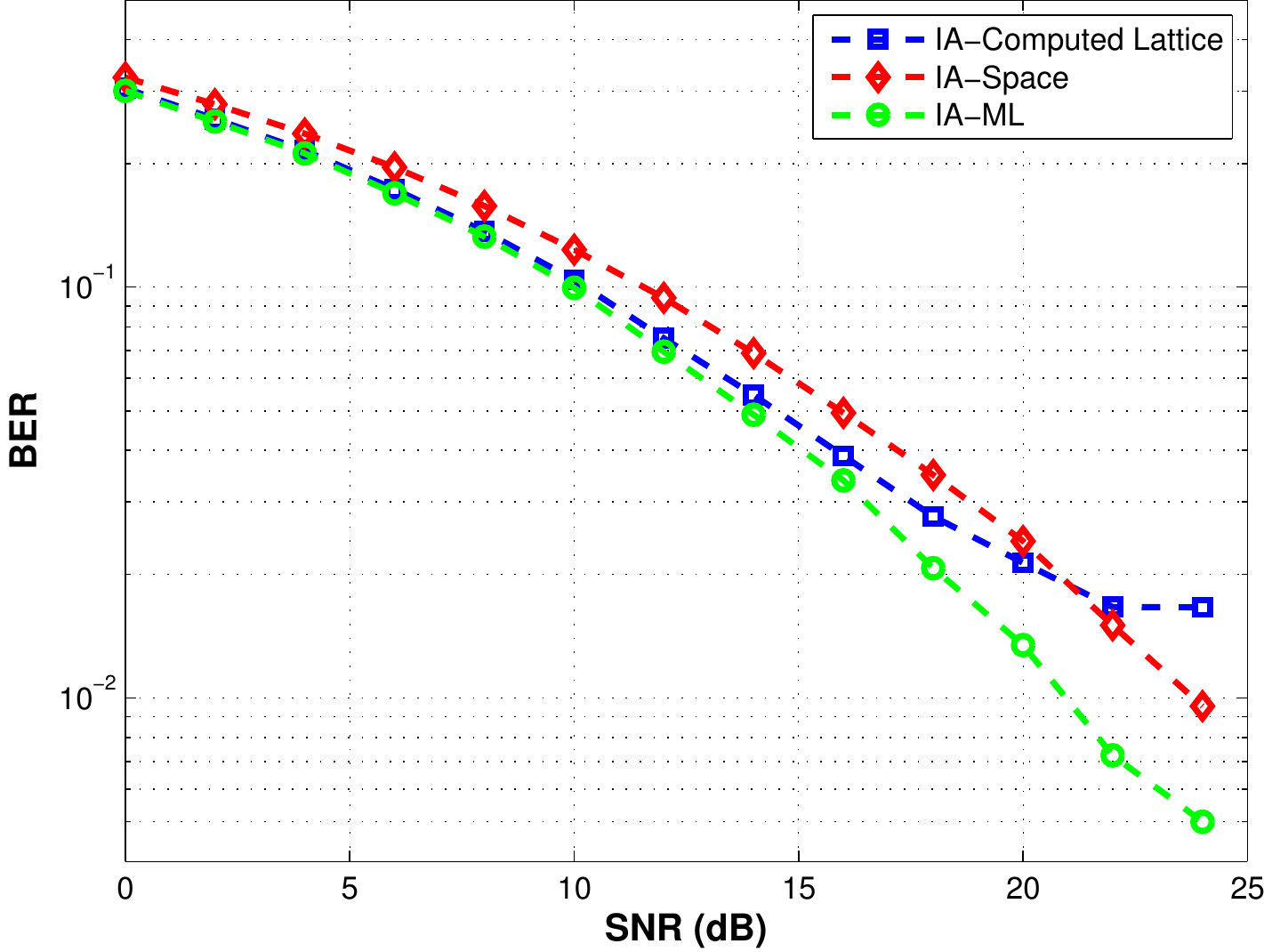}
	\label{figure:BER_QAM_ant_2}
			\vspace{-0.2in}
		\caption{BER Vs. SNR for the pseudo-lattice detector with S-IA precoders. Number of antennas $N$ = 2, 16QAM modulation, the channel variance $\sigma_h^2 = 1$ and the noise variance $\sigma_z^2 = 1$}
\end{figure}

\begin{figure}
	\centering
	\includegraphics[width=0.45\textwidth]{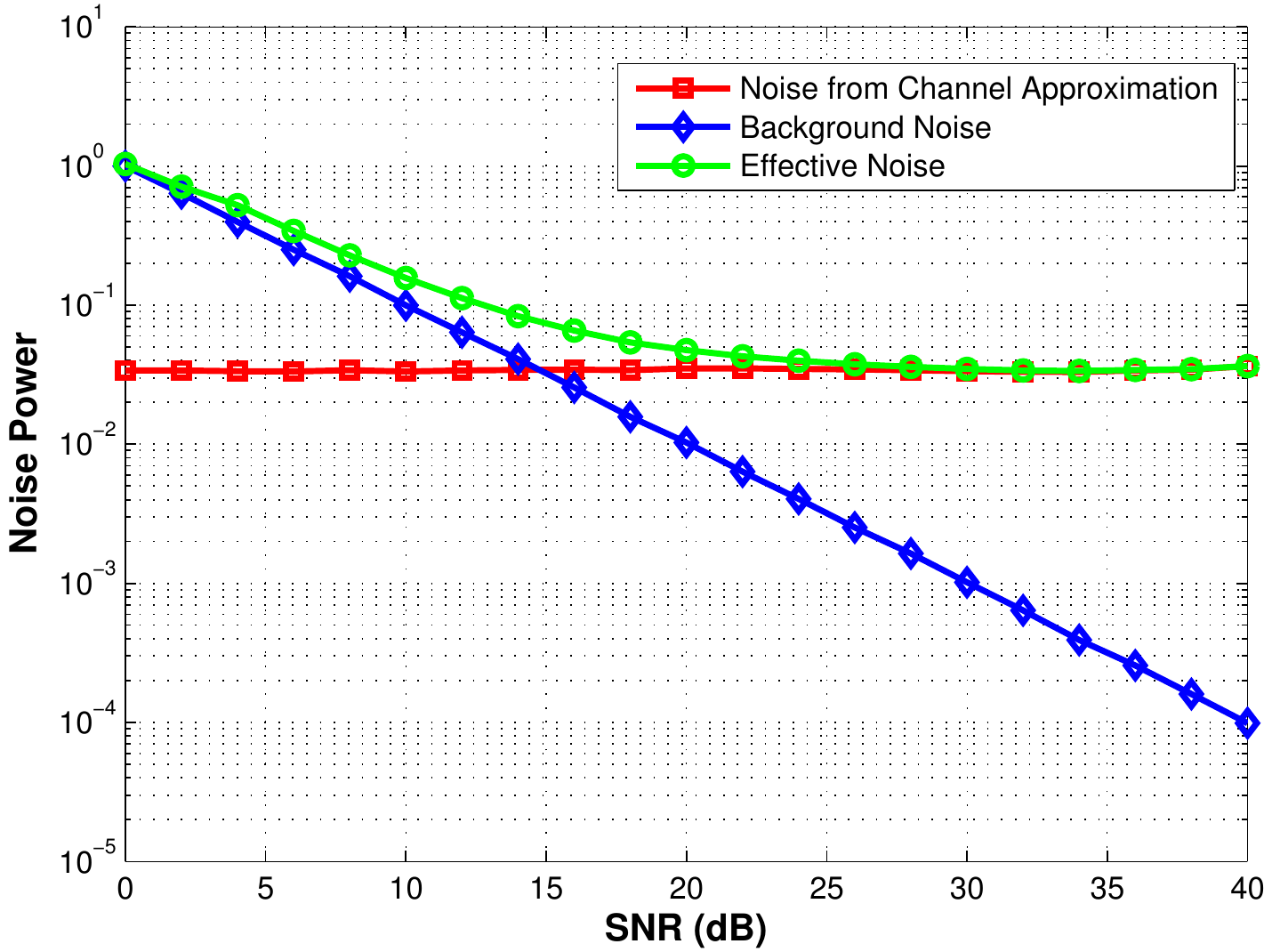}
				\vspace{-0.2in}
	\caption{Effective Noise at user 1 ($n_1$) Vs. SNR for the pseudo-lattice detector with S-IA precoders. $N = 4$, QPSK modulation. The channel variance $\sigma_h^2 = 1$ and the noise variance $\sigma_z^2 = 1$.}
	\label{figure:noise}
\vspace{-0.3in}
\end{figure}

Figure \ref{figure:noise}, shows the change of different component of the effective noise $\mathbf{n}_1$ with SNR. It shows from the figure that at low SNR conditions, the major noise source is the background noise. This can be shown by the effective noise tracing the background noise component in low SNR conditions. Since the channel approximation noise is dependent of the channel variances, the noise exhibits a flat energy value over different SNR conditions. In higher SNR,  background noise falls below the channel approximation noise and the effective noise power asymptotically traces the channel approximation noise. 

Figure \ref{figure:BER_QAM_ant_2} shows the simulation results for 16QAM constellation for $N=2$. Again we show the results for our proposed detector versus the ZF detector and the ML benchmark. Our proposed detector achieves the ML detector performance in the low and medium SNR vales. 

\section{Conclusion}
The use of Lattice Interference Alignment (L-IA) precoding enables the use of joint detection through LR-based detectors over MIMO multiuser interference networks. L-IA is a more strict approach to interference alignment than S-IA therefore it cannot be achieved through precoding for all nodes in the network. We therefore propose a pseudo-lattice detection which is LR-based to enable LR-based detectors benefits for these users. Our proposed technique can also be used for S-IA at all the receivers. We show that this approach can be implemented in a 3-user network for one of the users while not disrupting L-IA at the other two users. The performance of the detector is benchmarked against the ML-detector and compared to ZF detector for L-IA. An error floor in high SNR conditions is observed and studied for the proposed detector.
\bibliographystyle{IEEEtran}
\bibliography{interference_alignment}

\end{document}